\documentclass[11pt,hyper,letter]{JHEP3}
\usepackage{graphicx,amssymb,amsmath,amsfonts,cite}

\usepackage[english]{babel}
\usepackage{amssymb}
\usepackage{amsfonts}
\usepackage{amsmath}
\usepackage{graphicx}
\usepackage{epsfig}
\usepackage{cite}
\newcommand{\cA}{{\cal A}}  
\newcommand{\cC}{{\cal C}}  
  \newcommand{\cF}{{\cal F}}

  \newcommand{\cN}{{\cal N}}
\newcommand{\cO}{{\cal O}}

\newcommand{\be}{\begin{equation}} \newcommand{\ee}{\end{equation}}
\newcommand{\bea}{\begin{eqnarray}} \newcommand{\eea}{\end{eqnarray}}
\newcommand{\beann}{\begin{eqnarray*}}  \newcommand{\eeann}{\end{eqnarray*}}
\newcommand{\bfig}{\begin{figure}} \newcommand{\efig}{\end{figure}}
\newcommand{\ba}{\begin{array}} \newcommand{\ea}{\end{array}}
\newcommand{\bcen}{\begin{center}} \newcommand{\ecen}{\end{center}}
\newcommand{\btab}{\begin{tabular}} \newcommand{\etab}{\end{tabular}}

\newcommand{\bra}[1]{\langle #1|}
\newcommand{\ket}[1]{|#1\rangle}
\newcommand{\vev}[1]{\left\langle{#1}\right\rangle}

\newtheorem{Proposition}{Proposition}[section]

\newtheorem{Theorem}{Theorem}[section]
\newtheorem{Lemma}{Lemma}[section]

\newcommand{\bp}{\begin{Proposition}}   \newcommand{\ep}{\end{Proposition}}
\newcommand{\bt}{\begin{Theorem}}   \newcommand{\et}{\end{Theorem}}
\newcommand{\bl}{\begin{Lemma}}     \newcommand{\el}{\end{Lemma}}
\newcommand{\bc}{\begin{Corolary}} \newcommand{\ec}{\end{Corolary}}
\def\bra{\left\langle}
\def\ket{\right\rangle}
\def\vev#1{{\bra #1 \ket}}

\def\d{\partial}

\title{\LARGE Holographic Josephson Junctions and Berry holonomy  from D-branes}

\author{Sophia K. Domokos${}^1$\footnotemark[1], Carlos Hoyos${}^2$\footnotemark[2], Jacob Sonnenschein${}^2$\footnotemark[3]\,
\\
${}^1$\textit{Weizmann Institute of Science, Rehovot 76100, Israel}\\
\\
${}^2$ \textit{
  Raymond and Beverly Sackler Faculty of Exact Sciences \\
School of Physics and Astronomy \\
Tel-Aviv University, Ramat-Aviv 69978, Israel.}
}

\footnotetext[1]{E-mail address: \email{sophia.domokos@weizmann.ac.il}}
\footnotetext[2]{E-mail address: \email{choyos@post.tau.ac.il}}
\footnotetext[3]{E-mail address: \email{cobi@post.tau.ac.il}}

\abstract{We construct a holographic model for Josephson junctions
with a defect system of a $Dp$ brane intersecting a D(p+2) brane. In addition to providing a geometrical picture for the holographic dual, this leads us very naturally to suggest the possibility of non-Abelian Josephson junctions  characterized in terms of the topological properties of the branes. The difference between the locations of the endpoints of the $Dp$ brane on either side of the defect translates into the phase difference of the condensate in the Josephson junction. We also add a magnetic flux on the D(p+2) brane and allow it evolve adiabatically along a closed curve in the space of the magnetic flux, while generating a non-trivial Berry holonomy. }


\preprint{TAUP-2953/12\\ WIS/12/12-JULY-DPP}

\begin{document}

\section{Introduction}

The AdS/CFT correspondence provides a precise map between observables of a quantum field theory at strong coupling, and classical fields in a weakly coupled gravitational theory. Since the gravitational theory lives in a spacetime with  one dimension more than the field theory, the correspondence is also known as  a holographic duality. This map provides an obvious advantage for  computing quantities at strong coupling, and has motivated the study of many toy models aimed at applying the duality to real systems -- traditionally to QCD but also more recently to strongly correlated condensed matter systems (see \cite{CasalderreySolana:2011us,Kim:2012ey,Hartnoll:2009sz,Adams:2012th} for reviews on these topics).

One topic of interest for condensed matter applications of
holography, or AdS/CMT, is the realization of Josephson junctions
\cite{Josephson:1962zz}. These objects consist of two superconductors
separated by a ``weak link'' made of another material: an insulator
for ``SiS'' junctions, a normal metal for ``SnS'' junctions, or
simply a narrowing of the contact surface for  ``SsS'' junctions.
Generically, Josephson junctions describe interfaces in
superconductors across which the electron-pair condensate suffers a
change in phase. For SiS and SnS junctions, quantum mechanical
tunnelling across the interface induces a non-zero current
proportional to the sine of the phase difference, even in the
absence of applied voltage. This is the DC Josephson effect.
Meanwhile, if one applies a non-zero DC voltage across the junction,
one observes a current which oscillates in time (the AC Josephson
effect).

 A number of
works have recently used AdS/CMT techniques to model Josephson
junctions in various dimensions and configurations, for s-wave
\cite{Arean:2010xd,Horowitz:2011dz,Siani:2011uj,Wang:2011rva,Wang:2012yj}
and p-wave \cite{Wang:2011ri} superfluids, including a grid of
Josephson junctions in \cite{Kiritsis:2011zq}. These works are based
primarily on the Abelian-Higgs model for holographic superfluids,
with boundary conditions that break translational invariance. While
quite fruitful for numerical studies of the phase structure, these
models typically require one to solve complicated coupled partial
differential equations. The aim of this work is to construct a
simpler geometrical picture in terms of a D-brane model. In
particular we will present what to our knowledge is the first
explicit realization of a non-Abelian junction in holographic
models. In principle a non-Abelian Josephson effect could appear in
systems with an order parameter charged under a non-Abelian global
group, such as superfluid ${}^3 He$ \cite{Ambegaokar:1974}, high $T_c$
superconductors \cite{Demler:1998zz}, Bose-Einstein condensates of
atoms with spin \cite{Qi:2008} or the CFL (Color-Flavor-Locked)
superconducting phase of QCD \cite{Alford:2007xm}.

The theories we study are $(p+1)$-dimensional supersymmetric
$U(N_c)$ gauge theories at strong coupling, with a large value of
$N_c$, and correspond to the low-energy theory living on a stack of
$Dp$-branes. These theories have an $SO(9-p)$ global symmetry group,
that is partially broken when a few branes are separated from the
stack.\footnote{ The gauge group is
also partially Higgsed.} In this sense the system is dual to a
superfluid phase. The magnitude of the symmetry-breaking condensate
corresponds to the radial distance by which a small number of
$Dp$-branes are separated from the $\cO(N_c)$ stack. 

To this holographic $p$-dimensional superfluid, one can add a
co-dimension one defect by including a $D(p+2)$-brane intersecting
the $Dp$-branes. The defect explicitly breaks the global $SO(9-p)$
symmetry to a $SO(3)\times SO(6-p)$ subgroup, which is then
spontaneously broken to $SO(2)\times SO(5-p)$. When the $N_c\to
\infty$ limit is taken, the theory on the $Dp$-branes has a
holographic dual description, with probe $Dp$ and $D(p+2)$-branes in
a background geometry. These theories were first considered in a 
holographic context in \cite{Karch:2000gx}.

The defect is such that the (non-Abelian) phase of the condensate
can be different on either side of it. Therefore, the $Dp/D(p+2)$
intersection can be interpreted as a  holographic realization of a
Josephson junction. The effective description of the brane
construction is very similar to the field theory model of a
non-Abelian junction proposed in \cite{Esposito:2007fe}: the
fluctuations of the $Dp$ branes on both sides of the defect are the
Goldstone bosons of the spontaneously broken symmetry. In this way
the symmetry is naively doubled, but interactions between the fields
on either side break the symmetry to the diagonal subgroup, so one
is left with the correct number of true Goldstone bosons, while the other
modes become pseudo-Goldstones and are responsible for the Josephson
effect. In our case supersymmetry prevents the appearance of a DC
Josephson effect as we discuss in more detail below,  but additional
fluxes can be turned on in the $D(p+2)$ brane, that induce an AC
Josephson effect.

Though the $Dp/D(p+2)$ brane intersection is certainly an idealized
model, it nevertheless shares some interesting properties with the
physical examples mentioned above. In particular the ground state is
degenerate, meaning that there is a global symmetry that remains
unbroken in the superfluid phase. Let us assume that there are some
parameters that determine the properties of the junction, and that
these parameters can be changed adiabatically, in such a way that
the evolution forms a closed curve in parameter space. Since the
ground state is degenerate, for a non-Abelian junction the final
state of the system does not necessarily coincide with the initial
state: they may differ by a Berry holonomy \cite{Berry:1984jv}.

In the $Dp/D(p+2)$ intersection the parameters of the junction are
magnetic fluxes on the $D(p+2)$ brane, and the Berry holonomy
depends on the amount of electric flux, or equivalently, on the
number of $F1$ strings dissolved on the $D(p+2)$ brane. One can see
the Berry holonomy thus defined as a topological property of the
intersection, and it can be measured through the Josephson current
on the $Dp$ brane induced by the magnetic fluxes. Other definitions
of Berry holonomies in brane intersections are also possible: for example, they
have been studied in black hole systems corresponding to
D1/D5 intersections \cite{deBoer:2008qe} and for pairs of D0 branes
circulating around each other \cite{Pedder:2008je}.

The D3/D5 intersection deserves special attention as it was studied
in detail by Gaiotto and Witten in the context of supersymmetric
boundary conditions and $S$-duality
\cite{Gaiotto:2008sa,Gaiotto:2008sd,Gaiotto:2008ak}, including
fluxes on the $D5$ brane \cite{Gaiotto:2008sa}. Even with the string flux on the
D5, the system remains supersymmetric. The effect
of the flux can be seen as a modification of the boundary conditions
for the fields living on the $D3$ brane. Although this goes beyond the
scope of this paper, it would be interesting to see the relation
 to the Berry holonomy.

The paper is organized as follows: in Section~\ref{sec:defect} we
present the $Dp/D(p+2)$ intersections with electric flux and discuss
their holographic interpretation as models of supersymmetric superfluids
with a defect. In Section~\ref{sec:ac} we show how magnetic fluxes
on the $D(p+2)$ brane induce a Josephson current on the $Dp$
worldvolume in the Abelian case. We generalize to the non-Abelian
case in Section~\ref{sec:nonab}, and we compute the Berry holonomy
and connection in Section~\ref{sec:berry}. We end with the
conclusions and future directions of the results in
Section~\ref{sec:disc}.

\section{Supersymmetric superfluids with a defect as Josephson junctions}\label{sec:defect}

In this section we will present some simple models that describe
supersymmetric superfluids in different dimensions. The superfluids
have a co-dimension one defect where a Josephson effect can be
induced by changing some parameters. The effect is non-Abelian, and
it can be characterized by a Berry holonomy that we will define in
section \ref{sec:berry}.

Our starting point is the following D-brane configuration, consisting of a D3 and a D5 with $N_s$ units of electric flux $E$ on the D5 worldvolume:
\begin{equation}
\begin{array}{|c|c||cccccccccc|}
\hline \# & & 0 & 1 & 2 & 3 & 4 & 5 & 6 & 7 & 8 & 9  \\ \hline
 N_c & D3 & \bullet & \bullet & \bullet & \bullet & & & & & &   \\
 1 & D5 & \bullet & \bullet & \bullet & & \bullet & \bullet & \bullet & & & \\
  N_s & E & \bullet & & & & \bullet & & & & &\\ \hline
\end{array}
\end{equation}
There are $N_c$ $D3$ branes intersecting a $D5$ brane\footnote{The setup can
be generalized further by incorporating $N_f$ coinciding  $D5$ branes.} localized in
the $3$ direction. The D5 with electric flux can be interpreted as a bound state of a D5
brane with $N_s$ F1 strings. This is a 1/4 BPS configuration, as one can see
by taking T-dualities along the $1256$ directions, which results on
a configuration consisting of an $(N_s,1)$ string ending on a $D3$
brane.

One can easily generalize this construction for generic $Dp/D(p+2)$ intersections. The low-energy theory on the $D3$ branes is $3+1$ dimensional $\cN=4$
$U(N_c)$ super-Yang Mills. T-dualities along the $1$ and $2$
directions also allow one to construct lower dimensional supersymmetric
theories: a $2+1$ theory on $D2$ branes intersecting a $D4$ brane and a
$1+1$ dimensional theory on $D1$ branes intersecting a $D3$ brane.
One can also generate higher dimensional theories by taking
T-dualities on the $7$, $8$ and $9$ directions, giving $D4/D6$, $D5/D7$
and $D6/D8$ intersections.

The $Dp$ branes can end on the $D(p+2)$ brane, so the $Dp$ branes can be split in
two halves and separated along the directions parallel to the $D(p+2)$ brane. From the point of
 view of the fields of the $D(p+2)$ brane, the $Dp/D(p+2)$ intersections
 carry magnetic charges. From the point of view of the $Dp$, the intersection with the $D(p+2)$ brane is a codimension one defect.

We can construct a holographic dual of a superfluid phase for $p\leq 4$ as follows. We start taking the
number of $Dp$ branes $N_c$ to be very large. In this limit, one can substitute the $Dp$ branes by the geometry they source, and
the $Dp$ low-energy theory is described holographically by the near-horizon geometry \cite{Itzhaki:1998dd}:
\begin{equation}
ds^2=\left(\frac{R}{r}\right)^{(7-p)/2}dr^2+\left(\frac{r}{R}\right)^{(7-p)/2}\eta_{\mu\nu}dx^\mu dx^\nu+R^2\left(\frac{r}{R}\right)^{(p-3)/2}d\Omega_{8-p}^2.
\end{equation}
Here $d\Omega_{8-p}^2$ is the metric of a unit $(8-p)$-sphere, $S^{8-p}$. The ``conformal boundary'' is at $r\to \infty$. Except for $p=3$, there is a non-trivial dilaton profile
\begin{equation}
e^\phi=e^{\phi_0}\left(\frac{R}{r}\right)^{(7-p)(3-p)/4}.
\end{equation}
The only other background field that has non-trivial profile is an electric $(p+2)$-form flux on the directions
transverse to the sphere.\footnote{In the $p=3$ case the $F_5$ flux should be self-dual, so there is also flux on the $S^5$.} The potential $F_{p+2}=d C_{p+1}$ is
\begin{equation}
C_{p+1}=\frac{1}{g_s}\left(\frac{r}{R} \right)^{7-p}dr\wedge dx^0\wedge \cdots \wedge dx^p.
\end{equation}
The flux equals the rank of the group of the dual field theory or equivalently the number of $Dp$ branes.
The $D(p+2)$ brane becomes a probe in this geometry, wrapping a two sphere in the $S^{8-p}$, and extended along time, the radial direction and $p-1$ of the spatial directions $x^i$.

For $p=3$, the geometry is dual to $3+1$ dimensional ${\cal N}=4$ $SU(N_c)$ super Yang-Mills. One can obtain the lower dimensional theories through dimensional reduction.
The field content includes six real scalars $\Phi^\alpha$, $\alpha=1-6$ in the adjoint
representation whose eigenvalues parametrize the Coulomb moduli space that is locally $\mathbb{R}^6$. For instance, if one of the eigenvalues of one of the scalars has an expectation value,
\begin{equation}
\vev{\Phi^\alpha}=\delta^\alpha_1\left(
\begin{array}{ccc}
\vev{\phi_1} & & \\
& 0 & \\
& & \ddots
\end{array}
\right)
\end{equation}
the gauge group will be Higgsed $U(N)\to U(N-1)\times U(1)$ and there will be an spontaneous breaking of R-symmetry $SO(6)\to SO(5)$.\footnote{Since $\cN=4$ SYM is a conformal field theory, there is also a spontaneous breaking of conformal invariance.} Therefore, states on the Coulomb branch of $\cN=4$ can also be seen as (non-Abelian) superfluid phases, with $6 \times N$ possible order parameters (scalar condensates) that describe the breaking of both global and local symmetries. The story is similar for the lower dimensional theories, but the number of scalars increases
by the difference in the number of dimensions, so the R-symmetry group is $SO(7)$ in $2+1$ dimension and $SO(8)$ in $1+1$ dimensions.

In the brane picture, giving an expectation value to the scalars amounts to separating some $Dp$ branes from the stack. If this number is very small compared
with the total number of branes, the holographic dual description involves introducing a small number of $Dp$ branes probes in the dual geometry and
neglecting their backreaction. These branes are localized in the radial and the sphere directions, and extend along the $x^\mu$ coordinates. The position in the
radial direction is proportional to the absolute value of the condensate, while the position on the sphere fixes the breaking of global symmetry, as the original $SO(9-p)$ isometry of the sphere
is reduced to a smaller subgroup.

We call an idealized Josephson junction in three spatial dimensions a two-dimensional surface that connects two regions of space in which the condensate has a different phase.
In other words, the spontaneous symmetry breaking is different on either side of the junction, even though the absolute value of the condensate may be the same. In our language, it
should correspond to a defect where a $Dp$ brane changes its position in the sphere as it crosses it. Such
configurations can be realized by adding additional $D(p+2)$ branes in the way we have described before.
The intersection between the $Dp$ branes and the $D(p+2)$ is $(p-1)+1$ dimensional and will correspond to the surface of the Josephson junction.
The defect dual to the $D(p+2)$ brane carries its own degrees of freedom. The field theory on the defect has a global $U(1)$ symmetry,\footnote{This is for a single
$D(p+2)$. For $N_f$ $D(p+2)$ branes the symmetry is $U(N_f)$.} for which one can introduce a nonzero charge density. \footnote{Although the $Dp/D(p+2)$ brane configurations with electric flux we have described are supersymmetric, when they are introduced in the backreacted geometry of the $Dp$ branes they can potentially break supersymmetry and have an instability if they are dual to a theory with non-zero chemical potential. We want to thank Andreas Karch for pointing out this to us.}

The $Dp$ brane can end on the $D(p+2)$, so instead of having a
single $Dp$ brane intersecting the $D(p+2)$ brane, one can ``break''
the $Dp$ brane in two halves and move the endpoints of each half to
different positions on the $D(p+2)$. In this case, the endpoints can
move in the radial direction and in the directions of the $S^2$ that
the $D(p+2)$ wraps. In the first case the absolute value of
the condensate on each side of the defect is different, while in the
second case the pattern of symmetry breaking (the phase) is
different. In particular, if we move the $Dp$ branes along a maximal
circle in the $S^2$, leaving everything else fixed, the condensates
on both sides of the defect will differ only by an Abelian phase. If
we allow the endpoints to move anywhere on the $S^2$, the phase is
$SU(2)$-valued. We study the latter in section \ref{sec:berry}. The
configuration is illustrated in figure \ref{fig1}

\FIGURE[t!]{
\includegraphics[width=8.5cm]{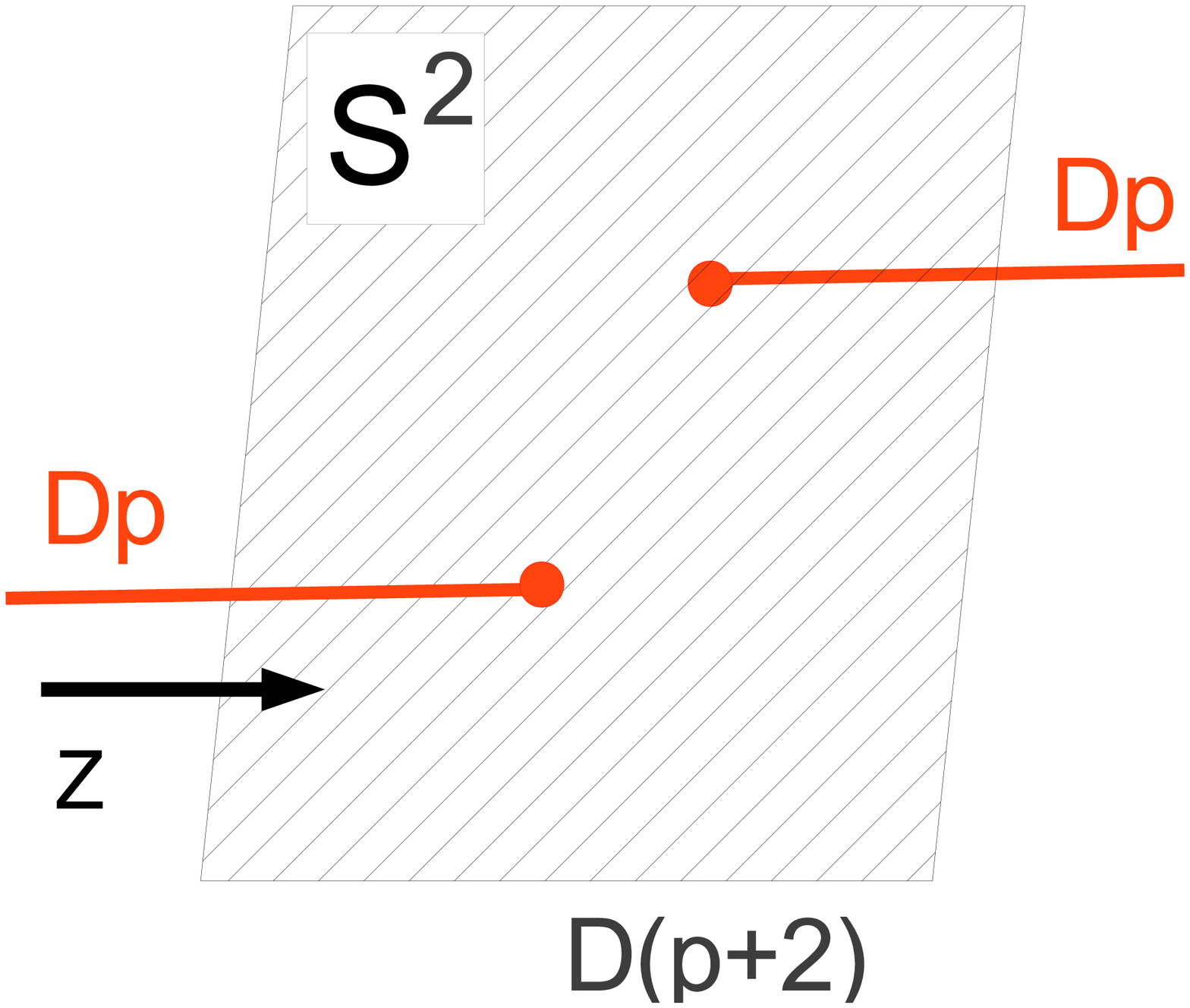}
    \caption{Holographic Josephson junction. A $D(p+2)$ brane is localized at $z=0$ and wrapping a $S^2$. $Dp$ branes end on the $D(p+2)$, that is seen as a codimension one defect in their worldvolume. The separation in the directions of the $S^2$ between the two $Dp$ halves corresponds to the jump in the phase of the condensate through the junction.}
     \label{fig1}
}

This gives a very simple and geometrical picture of a Josephson
junction.  Furthermore, the degeneracy of ground states makes this a
good scenario for studying non-Abelian effects.  However, because the
theory is supersymmetric, the system in this form does not admit a Josephson effect , since there is no force between the endpoints of the
$Dp$-branes. In order to observe a Josephson effect one needs to introduce fluxes on the $D(p+2)$ brane, as we will explain in the next section. In section \ref{sec:disc} we discuss other possible configurations that may show a Josephson effect even in the absence of fluxes.

\section{An AC Josephson effect from D-branes}\label{sec:ac}

We now show how to induce an AC Josephson effect in the $Dp/D(p+2)$
system. We will use the following coordinates for the
$(8-p)$-sphere
\begin{equation}
d\Omega_{8-p}^2=d\theta^2+\cos^2\theta d\Omega_{6-p}^2+\sin^2\theta d\phi^2 .
\end{equation}
The $D(p+2)$ brane wraps the coordinates $\phi$ and $\theta$, while
the endpoints of the $Dp$ branes are localized in all  directions on
the $(8-p)$-sphere. The directions along the $D(p+2)$ brane which
are transverse to the $Dp$ branes are $(\phi,\theta,r)$. One can
then see the endpoints of the $Dp$ branes as magnetic monopoles in a
3+1 dimensional spacetime.
The $Dp$ branes are flat along the $x^1,\cdots, x^{p-1}$ spatial directions but have a non-trivial profile in time and along the $x^p$ direction. Thus the only relevant coordinates that span the
worldvolume of the $Dp$ are $x^0= R t$ and $x^p= R z$. We will rescale the remaining spatial coordinates and the radial coordinate $r= R u$ by a factor of $R$ in such a way that they are dimensionless.  The profile
for each half-$Dp$ is given by
\begin{align}
X^0=t\quad X^\phi=\phi(t,z)\quad X^\theta=\frac{\pi}{2}+\theta(t,z)
\quad X^u=u_0+\delta u(t,z),
\end{align}
and the rest are zero or trivial.

The action for each half $Dp$ brane is then
\begin{align}
\notag S_{Dp} &=-\mu_p\int d^{p+1} x\,\left( e^{-\phi}\sqrt{-g}-C_{p+1}\right)\nonumber\\
\notag &=-T_pR^{p+1} u_0^{7-p}\int d^{p+1} x\left(\sqrt{1+u_0^{p-5}\left(\partial_\mu \phi\partial_\nu \phi+\partial_\mu \theta\partial_\nu \theta+u_0^{-2}\partial_\mu \delta u\partial_\nu \delta u\right)}-1 \right)\\
&\simeq -T_pR^{p+1} u_0^2\int d^{p+1} x\frac{1}{2}\eta^{\mu\nu}\left(\partial_\mu \phi\partial_\nu \phi+\partial_\mu \theta\partial_\nu \theta+u_0^{-2}\partial_\mu \delta u\partial_\nu \delta u \right)+\cdots,
\end{align}
where the $Dp$ brane tension is
\begin{equation}
T_p=\frac{\mu_p}{g_s}=\frac{1}{(2\pi)^p g_s(\alpha')^{(p+1)/2}}.
\end{equation}
Note that the  approximated action does not depend on the dimensionality of the
dual theory except in the overall constant factor. This should not
be surprising, as the interpretation of $\phi_a$ is that they are
the Goldstone bosons associated with  the $U(1)$ subgroup of the spontaneously broken
global symmetry of the dual field theory:  the free action for such
fields must take this form.
The equations of motion for the linearized fluctuations are just the Klein-Gordon equation in flat spacetime
\begin{equation}\label{eoms}
\square \phi =0, \ \ \square \theta =0, \ \ \square \delta u=0.
\end{equation}
The momentum densities are
\begin{equation}
\pi_\phi^\mu=-T_p R^{p+1} u_0^2 \partial^\mu\varphi, \ \ \pi_\theta^\mu=-T_p R^{p+1}  u_0^2 \partial^\mu\theta, \ \  \pi_u^\mu=-T_p R^{p+1} \partial^\mu \delta u.
\end{equation}
 For the rest of this section, we assume that the radial fluctuations $\delta u$ and the the fluctuations in $\theta$ vanish. The ends of the $Dp$-branes
 only fluctuate about their (antipodal) values $\phi=0$ and $\phi=\pi$. This is the `classic' Josephson effect,
 in which the amplitude of the condensate on  either side of the junction is equal, and the phase is Abelian.

Given that $\phi_a$ correspond to the phase of the condensate, the superfluid current is the conjugate canonical momentum
\begin{equation}
J_a^\mu=\pi_a^\mu=\frac{\delta S_{Dp}}{\delta(\partial_\mu\phi_a)}=-T_p R^{p+1}u_0^2\frac{\partial^\mu\phi}{\sqrt{1+u_0^{p-5}\eta^{\mu\nu}\partial_\mu \phi_a\partial_\nu \phi_a}}.
\end{equation}
From the perspective of the brane it corresponds to components of
its energy-momentum tensor. For instance, $\pi_a^0$ is the density
of momentum along the $\phi$ direction. In the absence of forces
$\pi_a^\mu$ is a conserved current, but if there is a force in the
$\phi$ direction $f_\phi$, then the momentum density current will
have non-vanishing divergence
\begin{equation}
\partial_\mu \pi_a^\mu=f_\phi^a,
\end{equation}
In particular, for homogeneous configurations
\begin{equation}\label{nonconsv}
\partial_0 \pi_a^0 = f_\phi^a \ \ \Rightarrow \ \ \partial_0 J_a^0=f_\phi^a,
\end{equation}
i.e. the time derivative of the superfluid charge density will be non-zero.

We can induce such a force by using the fact that, from the
perspective of the fields living on the $D(p+2)$ brane, the
endpoints are magnetically charged objects. They present a monopole anti-monopole system.  In a supersymmetric
configuration, the force induced by the bosonic fields on the
$D(p+2)$ cancels against the force induced by fermionic fields. In order to generate a non-zero force along the $\phi$ direction, we must introduce a
magnetic flux $F_\phi\propto B_\phi$, which corresponds to turning
on the $F_{\theta u}$ components of the field strength on the
$D(p+2)$ brane.\footnote{Note that this magnetic field is a flux in
the \textit{gravity dual} and is not a magnetic field in the
superfluid itself.}
In general this requires solving non-linear PDEs. We simplify the problem by assuming that the displacement of the $Dp$ branes around its equilibrium position is small, which is
a good approximation as long as the distance between the endpoints is not too large. This requires that the magnetic field on the brane is small and has an oscillatory behavior.
In principle we could also introduce a source in the $r$ direction by turning on the $F_{\theta \phi}$ components of the field strength on the $D(p+2)$ brane.
From the point of view of the $D(p+2)$, the endpoints of the $Dp$ are
magnetic monopoles. This should be analogous to  the problem of a string
ending on a D-brane (see for instance \cite{Zwiebach:2004tj}), except that the magnetic and electric fields
switch roles. Note that the $D(p+2)$ worldvolume is $(p+3)$-dimensional, so the magnetic dual
field strength
\begin{equation}
\cF={}^{*_{p+3}} F,
\end{equation}
is a $(p+1)$-form, and the magnetic dual potential a $p$-form. This potential has a natural coupling
to the intersection between the $Dp$ and the $D(p+2)$, that is $p$-dimensional.
Now we turn on the magnetic potential $\cA$ on the $D(p+2)$-brane. The action on the $Dp$ endpoints at $z=0$ should have the form
\begin{align}
S_{\cA}=\int_{(z=0)} d^p x \cA_{M_1M_2\cdots M_p}(X)\frac{\d X^{M_1}}{\d
t}\frac{\d X^{M_2}}{\d x^1}\cdots \frac{\d X^{M_p}}{\d x^{p-1}}~,
\end{align}
with $X^M$ the coordinates pullback of the scalar fields on the $Dp$ to the $Dp$-$D(p+2)$ intersection. We implicitly hide whatever factors of $u_0$ etc that arise into the normalization of $\cA_{M_1\cdots M_p}$. Since we want a uniform charge density in the $x^1$ to $x^{p-1}$ directions, we
only have nonzero components ${\cA}_{M12\cdots p-1}$ where $M$ can be
$t,u,\theta,\phi$.

Now let us vary the total action
\begin{align}
\delta \left( S_{Dp}+S_\cA \right)&= T_p\int d^{p+1}x \sqrt{-g}\delta X^M\Box
X_M\\
&+\oint d^p x\delta X^M\left[ -T_p\sqrt{-g}g_{MN}g^{zz}\d_z
X^N+\cF_{MM_1 M_2\cdots M_{p}}\frac{\d X^{M_1}}{\d t}\frac{\d X^{M_2}}{\d x^1}\cdots
\frac{\d X^{M_{p}}}{\d x^{p-1}} \right]_{z=0}\nonumber
\end{align}
Thus we see that the force on the $Dp$ endpoint acts as a modification
of the $z=0$ boundary condition so that
\begin{align}\label{bc}
T_p\sqrt{-g}g_{MN}g^{zz}\d_z X^N=\cF_{MM_1 M_2\cdots M_{p}}\frac{\d X^{M_1}}{\d t}\frac{\d X^{M_2}}{\d x^1}\cdots \frac{\d X^{M_{p}}}{\d x^{p-1}}.
\end{align}

For instance, if we turn on a magnetic potential of the form
$\cA_{\phi 12\cdots p}(X)=B\cos(\Omega X^0)$ we would have
\begin{align}
u_0^2T_pR^{p+1}\d_z \phi(z=0)=B\sin(\Omega t) ~.
\end{align}
and $\partial_z\theta=0$, $\partial_z u=0$. Electric fields would involve a coupling of the boundary
conditions for the coordinates $\phi$ and $\theta$, for example. The magnetic field is $B=\varepsilon B_0$, where $\varepsilon=\pm 1$ depending on the orientation of the endpoint.

There is an obvious solution to the equations of motion \eqref{eoms} with these boundary conditions:
\begin{equation}
u(t,z)=u_0, \ \ \theta(t,z)=\frac{\pi}{2}, \ \ \phi(t,z)=\phi_0-\frac{B}{u_0^2T_p R^{p+1} \Omega}\sin(\Omega (t-z)).
\end{equation}
This describes a right-moving wave on both $Dp$ branes. A left-moving solution is also possible.
The charge density transmitted through the junction  is
\begin{equation}
\dot{Q}=\left.\partial_0\pi^0_\phi\right|_{z=0}=-B\Omega \sin(\Omega t).
\end{equation}
If we add the change in both branes we see that the total variation vanishes. We can then compute the differential current density through the junction using charge conservation
\begin{equation}
\frac{d J^z}{ d z}=B\Omega \sin(\Omega t),
\end{equation}
If we assign a width $\ell$ to the junction\footnote{We introduce
this length for the purposes of aligning ourselves with the
condensed matter literature. In the framework of the model, the
junction has zero width -- or at most a width near the string
scale.}, then the current per unit area across the junction would be
\begin{equation}
J^z=-\dot{Q}\ell =B\Omega\ell \sin(\Omega t).
\end{equation}
This corresponds to an AC Josephson effect with amplitude $I_c=|B|\Omega\ell$ and a voltage across the junction $U=\frac{\Omega}{2}$.

We can also describe an approximate DC Josephson effect. First we introduce a magnetic field of the form
\begin{equation}
B_\phi=B \sin(\Omega t),
\end{equation}
 so that the charge transfer is
 \begin{equation}
\dot{Q}=-B\Omega\cos(\Omega t).
\end{equation}
Now we take the limit of small frequency and large amplitude $\Omega\to 0$  and $|B|\to \infty$ with $B\Omega =I_c$ fixed. Then, the charge transfer becomes approximately constant in time
\begin{equation}
\dot{Q}=-I_c\left[1+O((\Omega t)^2)\right].
\end{equation}
This would be a good approximation as long as $\Omega t\ll 1$.

\section{Non-Abelian Josephson junction}\label{sec:nonab}

So far we have studied a case where the difference between the condensates at both sides of the junction is just a phase, corresponding to the separation of the endpoints of the $Dp$ branes along a circle inside the $S^2$ wrapped by the $D(p+2)$ branes. In order to study non-Abelian effects we now will allow the endpoints of the $Dp$ brane to move on the full $S^2$:
\begin{align}
X^0=t\quad X^\phi=\phi(t,z)\quad X^\theta=\theta(t,z)
\quad X^u=u_0+\delta u(t,z),
\end{align}
where $\delta u$ will be taken to be small, but $\phi$ and $\theta$ can have variations of order one. We will work in an ``adiabatic'' approximation, meaning that we assume the derivatives of the fields to be very small,
though some of the field themselves can take on finite values. Then, to leading order, the DBI action on the $Dp$ worldvolume becomes
\begin{align}
S_{Dp}=-T_pR^{p+1}\int d^{p+1}\,x \frac{1}{2}\left[u_0^2\left( \sin^2\theta\d_\mu
\phi\d^\mu\phi+\d_\mu\theta\d^\mu\theta\right)+\d_\mu\delta
u\d^\mu\delta u \right].
\end{align}
The equation of motion and boundary conditions of $\delta u$ are unaffected in the new expansion, while the equations of motion of $\phi$ and $\theta$ are modified to
\begin{equation}
\partial_\mu \left[\sin^2\theta\partial^\mu\phi \right]=0, \ \ \square \theta-\sin\theta \cos\theta \d_\mu\phi \d^\mu\phi=0.
\end{equation}
These equations admit solutions that are a superposition of plane waves:
\begin{equation}
\theta(t,z) =\int \frac{d\omega}{2\pi}e^{i\omega(t-z)}\tilde\theta(\omega), \ \ \phi(t,z) =\int \frac{d\omega}{2\pi}e^{i\omega(t-z)}\tilde\phi(\omega).
\end{equation}
Therefore, one can introduce boundary conditions with arbitrary time dependence and
\begin{equation}\label{tequalsz}
\partial_z\theta(t,z=0)=-\partial_t\theta(t,z=0), \ \ \partial_z\phi(t,z=0)=-\partial_t\phi(t,z=0).
\end{equation}
The boundary conditions follow from the same analysis \eqref{bc}. Keeping only the leading terms in the derivative expansion we find:
\begin{align}\label{bc2}
\notag &u_0^2T_pR^{p+1}\sin^2\theta \d_z \phi =\cF_{\phi t12\cdots p-1}+\cF_{\phi \theta 12\cdots p-1}\partial_t\theta,\\
&u_0^2T_pR^{p+1}\d_z \theta =\cF_{\theta t12\cdots p-1}+\cF_{\theta\phi 12\cdots p-1}\partial_t\phi.
\end{align}
The first terms on the r.h.s. of the equations are the magnetic duals to magnetic fields on the $D(p+2)$ $F_{u\theta}$ and $F_{u\phi}$, while the last terms are dual to an electric field $F_{tu}$. We will introduce the following fluxes on the $D(p+2)$:
\begin{equation}
\cF_{\phi t12\cdots p-1}= B_\phi\sin(\Omega t), \ \ \cF_{\theta t12\cdots p-1}= B_\theta\cos(\Omega t), \ \ \cF_{\phi \theta 12\cdots p-1}=-\cF_{\theta\phi 12\cdots p-1}=E.
\end{equation}
In general $E$ depends on the $u$ coordinate, but the pullback to the endpoint is a constant. We assume that the frequency and the magnetic fields are all small, while the electric field is of order one. In terms of a small parameter $\epsilon$,
\begin{equation}
\d_z  \sim \d_t \sim B_\phi \sim B_\theta \sim \Omega \sim \epsilon, \ \ E \sim 1.
\end{equation}
This way all terms in the equations are of the same order. When \eqref{tequalsz} is satisfied, the boundary conditions become
\begin{align}\label{bc3}
\notag &-u_0^2T_pR^{p+1}\sin^2\theta \d_t \phi = B_\phi\sin(\Omega t)+E\partial_t\theta,\\
&-u_0^2T_pR^{p+1}\d_t \theta =B_\theta\cos(\Omega t)-E\partial_t\phi.
\end{align}
For simplicity we define rescaled time and fluxes
\begin{equation}
\tau=\frac{\Omega}{2\pi}t, \ \ e=\frac{E}{u_0^2T_pR^{p+1}}, \ \ b_\phi =\frac{2\pi B_\theta}{u_0^2T_pR^{p+1}\Omega}, \ \ b_\theta =\frac{2\pi B_\phi}{u_0^2T_pR^{p+1}\Omega},
\end{equation}
so the equations become
\begin{align}\label{bc3b}
\notag &-\sin^2\theta \d_\tau \phi = b_\phi\sin(2\pi\tau)+e\partial_\tau\theta,\\
&-\d_\tau \theta =b_\theta\cos(2\pi \tau)-e\partial_\tau\phi.
\end{align}
We can find a relation between the change in $\phi$ and in $\theta$ by integrating the last equation over a period
\begin{equation}
\Delta\theta=\int_0^{1} d\tau\,\d_\tau \theta=e\int_0^{1}
d\tau\,\d_\tau \phi=e\Delta \phi.
\end{equation}
Differentiating with respect to $\tau$ there are similar relations for higher derivatives, in general
\begin{equation}
\Delta(\d_\tau^n \theta)=e \Delta(\d_\tau^n\phi), \ \ n=0,1,\cdots.
\end{equation}

Solving for $\d_t\phi$ in the second equation, we get
\begin{equation}\label{solphi}
\d_\tau\phi=\frac{1}{e}\left(\d_\tau \theta+b_\theta\cos(2\pi\tau)\right).
\end{equation}
Plugging this expression in the first equation we are left with the first order equation:
\begin{equation}\label{eq:thetaeq}
(e^2+\sin^2\theta)\d_\tau\theta+b_\theta\cos(2\pi \tau)\sin^2\theta+ e b_\phi\sin(2\pi \tau)=0.
\end{equation}
The integration of  \eqref{solphi} leads to
\begin{equation}
\phi=\phi_0+\frac{1}{e}\left(\theta+\frac{b_\theta}{2\pi}\sin(2\pi \tau)\right).
\end{equation}
We cannot solve the equations analytically for general $e,b_\theta,b_\phi$, and resort to a numerical solutions.
An example of a trajectory on the $S^2$ is given in figure
\ref{fig:adiabatic}.

\FIGURE[ht]{
\includegraphics[width=8.5cm]{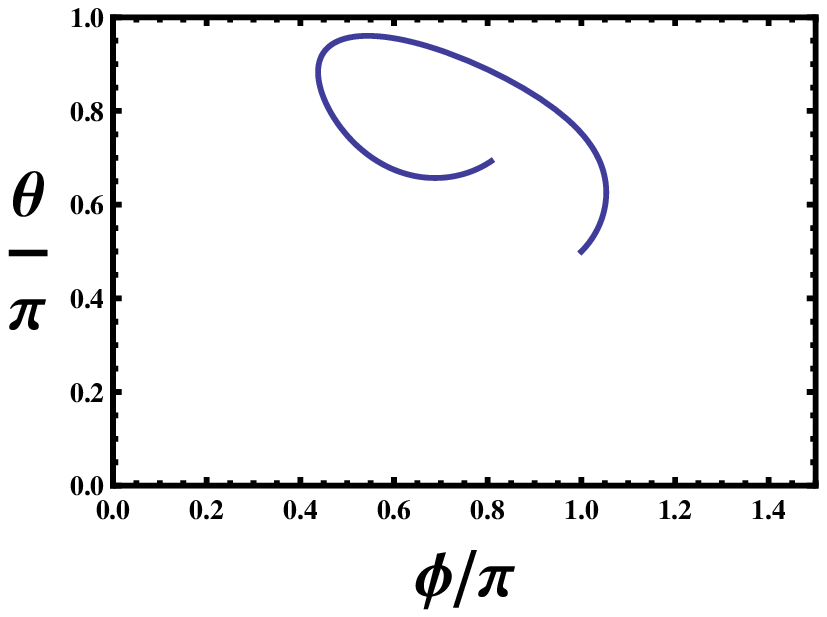}
\caption{\small
     The line represents the evolution of the solution in the $(\phi/\pi,\theta/\pi)$ plane, starting at $(1,1/2)$. After one period where the magnetic fields return to the same values, the value of the scalars is different. The values of the parameters for this solution are $e=-1$, $b_\theta=-5$ and $b_\phi=5$.
    }
    \label{fig:adiabatic}
}

When either of the magnetic fields vanishes, we can solve the equation \eqref{eq:thetaeq} analytically
(up to a transcendental relation). When $b_\theta=0$, we have
\begin{align}
&\left(e^2+\frac{1}{2}\right)\theta-\frac{1}{4}\sin
(2\theta)=-\frac{eb_\phi}{2\pi}\cos(2\pi\tau)+C_{0}\\
&\phi=\phi_0+\frac{1}{e}\theta=\phi_0+\frac{1}{e\left(e^2+\frac{1}{2}\right)}\left[
\frac{1}{4}\sin(2\theta
)-\frac{eb_\phi}{2\pi}\cos(2\pi\theta)+C_{0}\right]
\end{align}
where $C_{0}$ is the initial value of the transcendental,
\begin{align}
C_{ 0}=\left(e^2+\frac{1}{2}\right)\theta(\tau=0)-\frac{1}{4}\sin
(2\theta(\tau=0))+\frac{eb_\phi}{2\pi}.
\end{align}
The transcendental equation has only one solution
for each value of $\tau$. As one can see from the periodicity of the
solutions, then, in the case of $b_\theta=0$, both $\phi$ and
$\theta$ return to their original values after one cycle. The same
is true for $b_\phi=0$, where the solutions for the phase are
\begin{align}
&\theta-e^2\cos\theta=-\frac{b_\theta}{2\pi}\sin(2\pi\tau)-D_0\\
&\phi=\phi_0+\frac{1}{e}\left(D_0+e^2\cot\theta \right)
\end{align}
where
\begin{align}
D_0=\theta(\tau=0)-e^2\cos(\theta(\tau=0))~.
\end{align}
We can now see that in order to measure a non-trivial Berry holonomy
in the system, we must turn on $b_\phi$, $b_\theta$, and $e$.

\section{Berry holonomy and connection}\label{sec:berry}

In the holographic Josephson junction there is a degenerate space of configurations, that corresponds to moving the endpoints of the $Dp$ branes to different positions on the $S^2$ wrapped by the $D(p+2)$ brane. We have seen that as a consequence there is a non-Abelian Josephson effect and that the final state of an adiabatic evolution of the system along a closed curve in the space of magnetic fields is different from the initial state. We can make this statement more quantitative by defining a Berry connection in the space of magnetic fields and measuring the Berry holonomy along the closed path. As we will see, the evolution of the phase of the condensate and the Josephson current is determined by parallel transport along the curve with respect to the Berry connection.

Let us start mapping the two-sphere to the complex plane using a stereographic projection
\begin{equation}
z=\tan\frac{\theta}{2}e^{i\phi}.
\end{equation}
The group of automorphisms of the sphere is $GL(2,\mathbb{C})$, i.e. $2\times 2$ complex matrices with non-zero determinant:
\begin{equation}
g=\left(\begin{array}{cc}
a & b \\ c & d
\end{array}
 \right).
\end{equation}
The action of the group on the complex plane is
\begin{equation}\label{autom}
z'=\hat g z=\frac{a z+b}{c z+d}.
\end{equation}
We will actually be interested in a $SU(2)$ subgroup.

In the model of non-Abelian holographic Josephson junction we are studying, the magnetic fields are external parameters that we vary adiabatically. As the magnetic fields vary, the endpoints of the $Dp$ brane move on the two-sphere. This is seen as a change in the state of the system, that we can map to a trajectory in the complex plane through the stereographic map. This trajectory can be described using the group of automorphisms above. For instance, if the initial and final points of the trajectory in one period of oscillation are
\begin{equation}
z_i=\tan\frac{\theta}{2}, \ \
z_f=\tan\left(\frac{\theta+\beta_f}{2}\right)e^{i\alpha_f}.
\end{equation}
The two are related by the action of an element of an $SU(2)$ group
\begin{align}
\notag &z_f=\hat g(\alpha_f,\beta_f;0)z_i, \\ &g(\alpha_f,\beta_f;0)=g_\phi(\alpha_f)g_\theta(\beta_f)=\left(\begin{array}{cc}
e^{i\alpha_f/2} & 0 \\ 0 & e^{-i\alpha_f/2}
\end{array}
 \right) \left(\begin{array}{cc}
\cos \frac{\beta_f}{2} & \sin\frac{\beta_f}{2} \\ -\sin\frac{\beta_f}{2} & \cos \frac{\beta_f}{2}
\end{array}
 \right).
\end{align}
If the initial point is not real $z_i=\tan\frac{\theta}{2}e^{i\phi}$, then the group element relating the initial and final points is modified to
\begin{equation}\label{groupelement}
g(\alpha_f,\beta_f;\phi)=g_\phi(\alpha_f)g_\phi(\phi)g_\theta(\beta_f)g_\phi(-\phi)=\left(\begin{array}{cc}
e^{i\alpha_f/2} & 0 \\ 0 & e^{-i\alpha_f/2}
\end{array}
 \right) \left(\begin{array}{cc}
\cos \frac{\beta_f}{2} & \sin\frac{\beta_f}{2}e^{i\phi} \\ -\sin\frac{\beta_f}{2}e^{-i\phi} & \cos \frac{\beta_f}{2}
\end{array}
 \right).
\end{equation}

We can divide the trajectory into $N$ small pieces characterized by the time intervals $\Delta t=t_{i+1}-t_i$, $i=0,\cdots,N$. When they are glued together it is clear that the group element that relates the initial and final points is the product of group elements that relate the endpoints of each of the smaller intervals:
\begin{equation}
g(\alpha_f,\beta_f;\phi(0))=\prod_{i=0}^{N-1} g(\alpha(t_i),\beta(t_i);\phi(t_i)).
\end{equation}
We can make this an statement about an infinitesimal change along the trajectory. The change in the complex plane along the trajectory is
\begin{equation}
z(t+\Delta t)-z(t)=\partial_t z \Delta t=\frac{1}{2}\left(1+\tan^2\frac{\theta}{2}\right)e^{i\phi}\partial_t \theta\Delta t+i \tan\frac{\theta}{2}e^{i\phi}\partial_t \phi \Delta t.
\end{equation}
On the other hand, a transformation \eqref{autom} with infinitesimal values of $\alpha$ and $\beta$ in \eqref{groupelement} is
\begin{equation}
\hat g z-z =\frac{1}{2}\left(1+\tan^2\frac{\theta}{2}\right)e^{i\phi}\beta+i \tan\frac{\theta}{2}e^{i\phi}\alpha.
\end{equation}
Therefore, we can identify $\beta=\partial_t \theta\Delta t\equiv \dot \theta\Delta t$ and $\alpha=\partial_t\phi\Delta t\equiv \dot \phi\Delta t$. We can describe the trajectory using the equation
\begin{equation}\label{zdot}
\dot z= i\hat h(\delta a,\delta b,\delta c,\delta d) z,
\end{equation}
where $h$ is an element of the $gl(2,\mathbb{C})$ algebra, acting on $z$ as
\begin{equation}
\hat h(\delta a,\delta b,\delta c,\delta d) z=(\delta a-\delta d)z+\delta b-\delta c z^2.
\end{equation}
In our case it is an element of the $su(2)$ subalgebra. Although we do not know $h$ {\it a priori}, we can extract its value from the solutions we have found: 
\begin{equation}
h(t)=\dot \phi \frac{\sigma^3}{2}+\dot\theta\cos\phi\frac{\sigma^2}{2}+\dot \theta\sin\phi\frac{\sigma^1}{2},
\end{equation}
where $\sigma^i$ are the Pauli matrices.

In this way we can associate to each point of the trajectory an element of the $su(2)$ algebra. We can rewrite \eqref{zdot} as the equation of parallel transport along the trajectory in parameter space.
This allows us to define an $su(2)$ connection along the closed curve in the space of magnetic fields. We will identify this with the Berry connection along the curve. The unit vector tangent to the curve is
\begin{equation}
{\bf t}=\frac{1}{\sqrt{(\partial_t B_\phi)^2+(\partial_t B_\theta)^2}}(\partial_t B_\phi,\partial_t B_\theta).
\end{equation}
Then, equation \eqref{zdot} becomes
\begin{equation}\label{pt}
{\bf t}\cdot \left(\nabla -i \hat {\bf A}\right)z=0,
\end{equation}
where the value of the Berry connection along the curve is then defined as
\begin{equation}
{\bf t}\cdot \hat {\bf A}\equiv \frac{\hat h(t)}{\sqrt{(\partial_t B_\phi)^2+(\partial_t B_\theta)^2}}.
\end{equation}

Note that \eqref{pt} determines the parallel transport, as defined by the Berry connection, of the phase of the condensate along the curve in the space of magnetic fields. The Wilson loop along the closed curve in the space of magnetic fields determines the transformation \eqref{groupelement} that relates the two endpoints of the trajectory in the space of values of the condensate
\begin{equation}
g(\alpha_f,\beta_f,\phi(0))=e^{i\oint_{\cC} {\bf t}\cdot {\bf A}}.
\end{equation}
By Stokes' theorem, this should be related to the integral of the
Berry curvature in the area enclosed by the curve
\begin{equation}
\oint_{\cC} {\bf t}\cdot {\bf A} = \int_{\cA} F.
\end{equation}
In our examples the Wilson loop is non-trivial, indicating that there is a non-zero Berry curvature.

\section{Conclusions and future directions}\label{sec:disc}

In this paper we have used D-branes to construct a simple
holographic model for Josephson junctions in
$p+1$-dimensional superfluids, $0<p\leq 4$, as a $Dp/D(p+2)$ intersection. By varying the magnetic fields on the $D(p+2)$ brane we have found a non-trivial Berry holonomy, which measures the amount of electric flux, or equivalently the number of strings on the $D(p+2)$ brane. This should be an integer number, so the Berry holonomy must be quantized. It may thus be used as a way to characterize new topological phases of holographic superfluids.

For $p=3$ one could define a different kind of topological phase if one restricts to a particular class of models. The starting point is a single D5. By adding a large number of $NS5$-branes one can produce a supersymmetric Janus geometry where the theta angle on the $D3$ jumps across the defect, as suggested in
\cite{Gaiotto:2008sd}. In this case a Chern-Simons term for the $D3$
brane gauge fields is induced on the intersection between the $D3$ and
the $5$-branes. This implies that there is a Hall effect on the
defect, which one can see as the boundary between two systems.\footnote{A Hall effect can also be induced if there is a background axion and magnetic flux on the sphere that the D5 is wrapping\cite{Myers:2008me}.} This
is quite similar to the effective theory that describes the response
of topological insulators to external electromagnetic fields
\cite{TI,TSC}. In this sense one can see the coefficient of the
Chern-Simons term as labelling distinct ``topological phases'', its
value is quantized since the numbers of D5 and NS5 branes should be
integers.\footnote{This theory is not unique. In principle one can construct many supersymmetric Janus geometries where the theta angle is not quantized, but those would not correspond to a distribution of NS5 branes. We thank Andreas Karch for illuminating comments on this issue.} In the absence of $NS5$ branes there is no Chern-Simons term, so this would correspond to the ``trivial phase''. When the
fluxes on the $D5$ brane are turned on, the boundary conditions of the
fields on the $D3$ brane change. The $D3/D5$ intersection with string flux also
falls into the supersymmetric cases studied in \cite{Gaiotto:2008sa}, so it would be interesting to see the
connection of the boundary conditions with the Berry holonomy.\\

There are many other interesting directions which remain to be explored.\\

An obvious extension is to study intersections with more branes, $n_p$ $Dp$ and $n_{p+2}$ $D(p+2)$ branes. The relevant group of symmetries at the intersection is enlarged to $U(n_p)\times U(n_{p+2})\times SO(3)$, which can be broken in different ways as the $Dp$ branes are split, leading to different types of Josephson junctions. With several $D(p+2)$ branes it is also possible to build arrays of Josephson junctions, by separating the branes along the transverse spatial direction and connecting them with $Dp$ branes.

Another clear extension of this work would be to study other Josephson junctions obtained from $D$-brane configurations. A na\"\i ve candidate are $p$-dimensional intersections of $Dp$ and $D(p+4)$ branes. However, although in this configuration there is a codimension one defect on the $Dp$ branes, they cannot be split in two halves ending on the $D(p+4)$ branes. For $p=3$ there are no other possibilities left. For $p=1,2$ the remaining possibilities are $Dp/D(p+6)$ intersections $D1/D7$ and $D2/D8$, which are also supersymmetric.

Although the simplest models are supersymmetric, slightly more complicated models could lead to non-supersymmetric Josephson junctions, with a non-zero Josephson current even in the absence of additional fluxes on the defect. A possible candidate is the $(2+1)$-dimensional intersection $D3/D7' $ studied in  \cite{Bergman:2010gm} as a holographic model with a Quantum Hall Effect. Although the $D3$ cannot end on a $D7$, the $D7'$ has fluxes turned on that can be seen as $D5$ branes dissolved in its worldvolume, on which the $D3$ brane can in principle end.

Supersymmetry could also be broken explicitly, for instance by making one of the spatial
directions along the $D(p+2)$ compact with length $L$ and imposing
antiperiodic boundary conditions for the gaugino on the $D(p+2)$:
\begin{equation}
x^1\sim x^1+L, \ \ R\gg L \gg \sqrt{\alpha'}.
\end{equation}
If we choose the length of the compact direction to be much smaller
than the radius of the $S^{8-p}$, we can neglect the effect of the
massive gaugino on the $D(p+2)$. $L$ has still to be much larger
than the string scale in order to stay in the supergravity
approximation. This does not affect the stability of the $D(p+2)$
brane at the classical level, which is dependent only on the bosonic
fields. The final picture is that we put in contact two
superconductors of the same material but with different condensates
along a strip of width $L$.

One could also break supersymmetry by modifying the background geometry, for instance by introducing a temperature. This would also be interesting in this framework in order to be able to plot the phase diagram. However, if one naively replaces the $Dp$-brane background with a black hole,
the probe branes simply fall in, so the solution is not stable. One would therefore require a more complicated background, including additional bulk fluxes to counter the effect of gravity. Most likely in order to be able to pull $Dp$ branes outside the horizon it would be necessary for the background itself to be dual to a superfluid phase, so the phase diagram will be determined by the background. There are several models constructed from consistent truncations of supergravity (e.g. \cite{Gubser:2009qm,Arean:2010wu,Aprile:2010ge,Aprile:2011uq,Donos:2011ut}), although in order to introduce $D$-branes one should uplift them to ten dimensions.

Supergravity backgrounds can also be useful to go beyond the probe approximation. An approach directly related to the D-brane constructions we have presented will be to replace the defect brane by a Janus geometry \cite{Bak:2003jk,Clark:2004sb}, while keeping the $Dp$ branes as probes. For the intersection of $D3$ branes with $5$-branes there are known solutions \cite{D'Hoker:2007xy,Gaiotto:2008sd} which include
configurations with the $5$-branes separated in different stacks \cite{D'Hoker:2007xz,Aharony:2011yc}.

\section*{Ackwnoledgements}

We thank Andreas Karch for very useful comments and discussions on the manuscript.  This work was supported in part by the Israel Science Foundation (grant number 1468/06).

\end{document}